\documentstyle[preprint,aps]{revtex} 
\renewcommand{\baselinestretch}{1}

\begin{document}

\preprint{\vbox{\hspace*{\fill} DOE/ER/40762-112\\
          \hspace*{\fill} U. of MD PP \#97-080}} \vspace{.5in}

%JG: Title change made by PRC is only difference from "prl9770220PRppsub.latex"
\title{APEX: Positive evidence for sharp 800 keV pairs from heavy ion
collisions near the Coulomb barrier}

\author{James J. Griffin*}

\address{Department of 
Physics, University of~Maryland, College~Park, MD~20742}

\maketitle
\widetext
\begin{abstract} 
The best fit to the published APEX U+Th data describes a sharp line of
123 pairs of width 23 keV at 793$\pm$7 keV; also, the data impose a
positive 99.0\%CL {\it lower} bound of 23 sharp pairs.  It is therefore
untenable to argue from the APEX data against the existence of sharp
pairs. Data-only ratios APEX/EPOS pair counts show empirically that the
two experiments' pair data are mutually consistent and of comparable
statistical potency: conflicts, if any, can be resolved only by other
independent evidence. A perspective is offered on the current status of
the Sharp Lepton Problem, and alternative non-heavy-ion experimental
approaches to it are recommended.
\end{abstract}

\vspace*{0.5in}
{*Email: griffin@quark.umd.edu; Tel: 301-405-6115; FAX: 301-405-6114}\\

\vspace*{\fill}
\noindent {submitted to Phys. Rev. C; archived as paper nucl-th/9703006.}
\noindent{Typeset usingREVTEX}
%\pacs{25.70.Bc, 23.20.Nx, 21.45.+v, 14.80.-j, 14.60.Cd, 12.20.-m.}

\renewcommand{\baselinestretch}{1}
\narrowtext
\section{Introduction}
 
The APEX experiment was mounted to test EPOS' earlier observations of
sharp $(e^+e^-)$ pairs from high-Z heavy ion collisions. In a recent
letter the APEX collaboration\cite{ahma/95} presents its results, and
considers the earlier EPOS U+Th
results\cite{cowa/86,cowa/87,cowa/88a,sala/90} in the light of their
own new data. It asserts that their data offers no statistically
significant evidence of sharp pairs and concludes ``that the results of
the present experiment represent a real disagreement with the previous
observations''.

Since the present report draws a seemingly contrary conclusion from the
very same data, it is essential to stipulate at the outset that the
APEX (U+Th) data does unquestionably  exclude an excess sharp pair
count near 800 keV as large as the $\approx$2560  counts which they
were expecting. In fact the APEX report establishes a specific (99\%CL)
upper bound\footnote{APEX presents its upper bounds as cross sections.
We prefer to present the discussion in terms of counts, in order to
keep it independent of assumptions about the energy dependence of the
sharp pair production cross section.  We use the conversion factor,
2560/5=512 excess counts/($\mu$b/sr), to convert from APEX' sharp pair
production cross section to sharp pair counts. See also footnotes No. 7
and No. 26.} of 292 excess counts near 800 keV.  That discrepancy leads
the APEX collaboration to question the existence of the previously
reported sharp pairs, and in particular of the EPOS line near 800 keV,
and to draw the conclusion quoted above, of a ``real disagreement''
with the EPOS results.

The present analysis\footnote{Brief summaries of the present work have been presented in Refs.\cite{grif/97,grif/97u}.} confirms these results, and prescribes a range of
even smaller upper bounds upon the excess pair counts near 800 keV:
from 217 to 252 counts.  Nevertheless,  the present analysis also
provides statistically significant positive evidence that APEX actually
counted 123$\pm$46 sharp pairs near 800 keV.  This inference
contradicts none of APEX empirical evidence. In particular it honors the
quantitative upper bound of the APEX report.
In the end,  we find that the APEX U+Th experiment recorded (at
793$\pm$7 keV) 123$\pm$46 sharp pairs, above a background of 1480
pairs/20keV, among a total of 40.8K EPOS-type\footnote{``EPOS'-type''
pairs are those which are Qualitatively Similar to the pairs accepted
by the EPOS apparatus, as discussed in more detail in Section V.A
below.} pairs.  The EPOS experiment recorded (at 809$\pm$8 keV)
97$\pm$38 sharp pairs\footnote{This is the count given in
Ref.\cite{cowa/96} of excess pairs above the full (five run) EPOS
background without the Time of Flight selection. It may be compared
with the 105$\pm$20 counts of Ref\cite{sala/90} extracted from the two
highest energy runs alone. Without the EPOS Time-of-Flight gating, the
latter count increases  to 113$\pm$28, according page 156 of
Ref.\cite{sala/90}.} above a background of 1280 pairs/20keV among a
total of 50K pairs. Thus under direct comparison, the APEX and the EPOS
pair data sets are of comparable size, and yield comparable numbers of
background pairs and of sharp pairs above background near 800 keV.
There is therefore no significant experimental evidence of
contradiction between them.  It follows also that as regards coincident
pairs neither experiment can claim clear statistical superiority for
resolving any differences in their pair distributions.

We note that we consider these pairs to be of very special interest
because they
suggest\cite{grif/89ae,grif/91a,grif/93a,grif/94,grif/95e,grif/96se}
the unexpected existence of a tightly bound Quadronium
$(e^+e^+e^-e^-)\;$ particle, $Q_o$, which might affect\cite{grif/95e} the
discrepant 3$\gamma$ decay of positronium, and the
scattering\cite{grif/96se} of few MeV electrons, positrons, and of
$\sim$1.8 MeV photons, from high-Z nuclei, as discussed further in
Section VI.

\section{APEX' Published Pair Sum Energy Data}

Figure 1 summarizes the published\footnote{The author is grateful to
the APEX collaboration for supplying the data published in Figure 2 of
Ref\cite{ahma/95} in numerical form.} APEX data on 22K $(e^+e^-)$
pairs\footnote{These are pairs selected by APEX to resemble those
accepted by the EPOS experiment: The positron and the electron must be
observed in opposite arms of the detector, (``RL'' pairs),  and the
lepton energies must fall into a ``wedge cut'' similar to (but not
identical with) that imposed by EPOS to enhance its sharp pair signal.
We note however that these APEX pairs are not``Qualitatively Similar''
EPOS-type pairs (in the sense discussed in Section V.A), because they
include ``(1,n)'' pairs from events in which any  number, n, of
electrons may be detected in coincidence with the positron, whereas
EPOS accepted only ``(1,1)'' pairs from events in which one and only
one electron is coincident with the positron.} from its U+Th study with
sum energies in the range from 200 to 1400 keV. It presents the number
of counts observed in each 20 keV bin, and compares it with APEX'
background distribution of pairs which results from random
electron-positron coincidences.  The shape of the latter
``event-mixed'' distribution was measured by APEX by taking a random
positron from one event together with an electron from the subsequent
event\footnote{The true smooth pair background also includes pairs
from  nuclear IPC processes\cite{cowa/88a,sala/90}. Moreover, if
Quadronium decay occurs\cite{grif/95e,grif/96se}, background pairs (as
well as energetic photons easily mistaken for nuclear gamma rays) may
also arise from its most probable decay to $(e^+,e^-,\gamma)$.  For
these and other reasons, the present event-mixed background is subject
to an unknown systematic error not included in the analysis.
Nevertheless, for simplicity, and to register our results most directly
with those of the APEX report, we here take the APEX event-mixed
distribution of Table I, without adjustment, to describe the smooth
pair background in the region of interest.}. Also shown is the APEX'
simulated sharp pair distribution, N$_{SSP}$,  tabulated in Table I,
reduced by a factor of 10 from the 5$\mu$b/sr value expected by APEX.
All of the data of Figure 1 were presented graphically  in Figure 2a of
the APEX report, and in Figs.  VI.1.2, and VI.1.3 of the thesis of
M.R.  Wolanski\cite{wola/95} on the APEX U+Th experiment.

\subsection{Sharp Pairs Near 800 keV}

The present analysis focusses upon pairs in the energy range from 500
to 900 keV, where earlier
studies\cite{cowa/86,sala/90,berd/88,koen/89,koen/93} have indicated
the occurrence of pairs of sharp sum energy.  Table I presents the
relevant APEX U+Th data in numerical form.  It lists twenty-one sum
energy bins from 510 to 910 keV,  and their observed pair counts,
$N_{OBS}$.  Columns three through seven present the APEX' event-mixed
background distribution, $N_{BK}$, its simulated\footnote{This is the
distribution plotted in Figure 1 of the 2914 counts computed by APEX'
simulation program to occur if a 1.822
MeV source of sharp pairs were produced
with an average cross section of 5$\mu$b/sr, and decayed at rest in the
C.of M. frame. Our gaussian best fit to it has 2560 counts, a width of 18.9
keV, and a maximum at 804.3 keV.  See also footnote No.1.} sharp pair
distribution, $N_{SSP}$, the number of ``Excess'' (above the
background) sharp pairs, $N_{EXC}$, the values of $\chi_i^{APX}=
(N_{OBS}-N_{BK})/(N_{OBS})^{1/2}$, and the APEX inferred upper bound upon the 
pair production cross section.  Columns eight and nine list the
values of our best fitting ``One Sharp Line'' distribution, $N_{FIT}$,
and of its  deviations, $\chi_i$, from $N_{EXC}$ in units of the
standard deviation, $\sigma = ({N_{OBS})^{1/2}}$, and columns ten  and
eleven list the 99\% CL lower and upper bounds upon the mean sharp pair count
in each bin, calculated as described below.

Table I also  provides the chi-squared sum for testing this APEX data
against the hypothesis that it consists only of APEX' event-mixed
background. For these twenty one bins\footnote{For the sixty 20 keV
bins (with energies 210 keV to 1390 keV) presented in Figure 1, the
corresponding $\chi^2_{59}$ value is 65.76, yielding a reduced
chi-squared value of 1.11. The APEX' background therefore serves better
to describe APEX pair data over the  full 210-1290 keV APEX energy
range than over the 510-910 keV range considered here.} the value of
chi-squared is 28.30, so that the reduced chi-squared value for 20
degrees of freedom has an acceptable value of 1.41.  Although this
Background-Only assumption  provides a moderately good fit to this
data, we shall see that the assumption of one sharp line yields a
substantially better fit, tabulated in columns eight and nine.

Figure 2 plots in the 500-900 keV range of Figure 1 the observed  pair
count, $N_{OBS}$, measured by APEX (from column 2 of Table I).  It also
exhibits as ``error bars'', the standard deviations,
$\pm\sigma=(N_{OBS})^{1/2}$, which measure the fluctuations expected in
each bin count due to Poisson statistics alone. Also plotted in Figure
2 is the APEX' event-mixed background (solid line)  and the added
(cross hatched) contribution of the best fitting, ``One-Sharp-Line''
distribution, from columns three and eight of Table I, respectively.
(The fitting process is detailed below.) 

\widetext
\begin{center}
\renewcommand{\arraystretch}{1.1}
\begin{tabular}{|c|lcccc|c||cc||cc|}  \hline 
\multicolumn{11}{|c|}{TABLE I: U+Th PAIR DATA, FITS and BOUNDS}\\ \hline
\multicolumn{7}{|c||}{APEX' RESULTS:} &\multicolumn{2}{|c||}{One Line} &\multicolumn{2}{|c|}{99\%CL }\\ 
\multicolumn{7}{|c||}{Measured, Simulated and Inferred} &\multicolumn{2}{|c||}{Best Fit} &\multicolumn{2}{|c|}{Bounds}\\ \hline \hline
$E_{SUM}$ & $N_{OBS}$\hspace{2mm}    &$N_{BK}$\hspace{3mm}  & $N_{SSP}$\hspace{3mm} &$N_{EXC}$\hspace{2mm}
& $(\chi_i)^{APX}$ & (d$\sigma/d\Omega)^{UB}$&N$_{FIT}$ & $\chi_i$ &$\nu_L$ &$\nu_U$\\ \hline
   510 & 610 & 608.7 & 5.3    &+1.3 &+0.1&0.80  &0.0  &+0.1 &0.3  &66.7\\ 
   530 & 634 & 629.0 & 4.0   & +5.0 &+0.2&0.72  &0.0  &+0.2 &0.4  &70.6\\ 
   550 & 653 & 641.2 & 2.6   &+11.8 &+0.5&0.68  &0.0  &+0.5 &0.5  &76.6\\ 
   570 & 635 & 647.3 & 2.6  &-12.3  &-0.5&0.45 &0.0  &-0.5  &0.2  &59.0\\ 
   590 & 640 & 646.7 & 1.3  & -6.7  &-0.3&0.44 &0.0  &-0.3  &0.3  &62.8\\ 
   610 & 607 & 643.1 & 4.0 & -36.1  &-1.5&0.28 &0.0  &-1.5  &0.1  &45.4\\ 
   630 & 627 & 645.8 & 5.3  &-18.8  &-0.7&0.32 &0.0  &-0.7  &0.2  &54.8\\ 
   650 & 591 & 637.4 & 21.2 &-46.4  &-1.9&0.21 &0.0  &-1.9  &0.1  &40.7\\ 
   670 & 657 & 630.1 & 30.5 &-26.9  &+1.0&0.47  &0.0  &+1.0 &0.9  &89.8\\  
   690 & 650 & 628.1 & 22.5 &-21.9  &+0.9&0.42  &0.0  &+0.9 &0.7  &85.0\\  
   710 & 608 & 613.8 & 50.3 & -5.8  &-0.2&0.28  &0.0  &-0.2  &0.3  &61.8\\ 
   730 & 616 & 618.8 & 62.2  &-2.8  &-0.1&0.28  &0.0  &-0.1  &0.3  &64.2\\ 
   750 & 621 & 607.7 & 55.6  &+13.3 &+0.5&0.32  &0.1  &+0.5  &0.5  &76.2\\ 
   770 & 614 & 602.1 & 95.3  & +11.9  &+0.5&0.30  &11.8  &0.0 &0.5 &74.8\\  
   790 & 682 & 599.0 & 748.1  & +83.0 &+3.2&0.57  &83.1  &0.0 &24.1 &145.4\\ 
   810 & 613 & 584.5 & 1742.4  &+28.5 &+1.2&0.33  &28.2  &0.0 &1.0 &89.1\\ 
   830 & 567 & 574.8 & 62.2   &-7.8   &-0.3&0.21  &0.3   &-0.3 &0.2 &58.2\\ 
   850 & 512 & 561.5 & 4.0  &-49.5   &-2.2&0.12   &0.00  &-2.2 &0.1 &35.9\\ 
   870 & 583 & 546.2 & 0.   &+36.8   &+1.5&0.32   &0.00  &+1.5 &1.7 &95.2\\  
   890 & 517 & 529.8 & 0.  &-12.8   &-0.6&0.17   &0.00  &-0.6 &0.2  &52.4\\ 
   910 & 502 & 504.9 & 0.  & -2.9   &-0.1&0.18   &0.00  &-0.1 &0.3  &57.9\\ \hline \hline
\multicolumn{7}{|c||}{$(\chi^2_{20})^{APX}=\Sigma(\chi^{APX}_i)^2 = 28.30$}&\multicolumn{4}{|c|}{$\chi_{17}^2 = \Sigma(\chi^2_i) = 16.65$}\\
\multicolumn{7}{|c||}{$(\tilde{\chi}^2_{20})^{APX} = 1.41$}&\multicolumn{4}{|c|}{$\tilde{\chi}_{17}^2 = 0.98$}\\ \hline
\end{tabular}
\end{center}

\widetext
\footnotesize
\begin{flushleft}
 Table I. The table lists the APEX pair data in the range 500-920 keV,
 from Figure 2a of Ref.\cite{ahma/95}, as follows. (1)The bin energies,
 $E_{SUM}$,  (2) APEX' pair counts, N$_{OBS}$, (3) APEX' event-mixed pair
background, N$_{BK}$,and (4) N$_{SSP}$, APEX' simulation calculation of
the sharp pair distribution from back-to-back sharp pair decay at rest
in the C.  of M. frame of an object produced with a cross section of
5$\mu$b/sr. Column five presents the excess pair count above the
background, N$_{EXC}$, column six the values of  $\chi_i^{APX}$,
whose squares sum to $(\chi_{20}^2)^{APX}$, and column seven, APEX' upper
bound (in $\mu$b/sr)  on the pair cross cross section. Columns eight and nine
provide our best Gaussian fit, N$_{FIT}$, to the excess in column 5,
and the remnant deviations, $\chi_i$, of that best fit from the
observed values, whose squares sum to $\chi_{17}^2$. The last two
columns list $\nu_L$ and $\nu_U$, the lower and upper bounds upon the
mean excess counts in each bin implied (at the 99\% confidence level)
by the observed counts.
\end{flushleft}
 
\normalsize

\narrowtext

It is evident in Figure 2, that the largest excess pair count occurs in
the bin centered at 790  keV, and that its magnitude (83.0 counts) is
about 3.2$\sigma$.  Such a large excess ought statistically to occur
only once in 500 such one-bin measurements\footnote{Since the APEX
experiment consists of some 60 such single-bin measurements, one
statistical fluctuation of this relative magnitude in an APEX
experiment is a mildly improbable event.}. Figure 2 requires that any
attempt to argue on the basis of  the APEX experiment against the
occurrence of sharp pairs near 800 keV must address the significance
of   this 790 keV excess.

\section{Chi-Squared One-Sharp-Line Analysis of  APEX' Pair Data}

In fact, the assumption that a gaussian-shaped distribution of sharp
pairs describes these excess counts allows a precise fit to this
excursion, and effects a statistically significant reduction in the
value of chi-squared. Therefore we compare with APEX'
``Background-Only'' hypothesis  the alternative ``One-Sharp-Line''
hypothesis that the excess pair count is well explained by the sum,
$N_{FIT}^{TOT} = N_{BK} + N_{FIT}$, of the event-mixed background and a
distribution, $N_{FIT}$, arising from a single sharp pair line.
Specifically, we define

\begin{equation}
 N_{FIT}(E_i) = \int_{\Delta E_i}D(E; S,\Gamma,E_S)dE,
\end{equation}

\noindent  to specify the number of counts from the sharp line
distribution which fall into the bin, $\Delta E_i$, near $E_i$, and take
the sharp  pair energy distribution, D, to be of  gaussian
form centered at energy $E_S$, with FWHM of $\Gamma$, and with 
an integrated total number of pairs equal to S. These three
parameters are chosen to minimize the chi-squared sum over all the
bins, to which each bin contributes a term,

\begin{equation}
(\chi_i)^2 = (N_{OBS} - N^{FIT}_{TOT})^2/(N_{OBS}),
\end{equation}

\noindent where the values, $\chi_i$, are listed in Table I (as are the corresponding values, $\chi_i^{APX}$, for the APEX' Background-Only fit).

\subsection{ Chi-Squared Results for APEX' Sharp Pairs}

The quantitative numerical results of the chi-squared analysis are
summarized in Table II and Figure 3. One finds that the minimum value of
chi-squared occurs at the values\footnote{We have also executed a four
parameter (B,$\Gamma$,E$_S$,S) fit in which the coefficient, B, of the
background term is allowed to vary, together with the three parameters
of the sharp line. One finds the best fitting values (B= 0.991, 25.7,
792.4, 140.6) and $\chi^2$=15.70 $(\tilde{\chi}^2=0.981)$, as compared
with the vaues (B$\equiv$1.0, 23.1, 792.8, 123.4) and $\chi^2$= 16.65
$(\tilde{\chi}^2=0.979)$ obtained here by keeping the background
coefficient fixed at the APEX' value. Since the present
fixed-background B$\equiv 1$ fit yields a smaller value for the total
strength of the sharp pair line, it understates the case for the
occurence of such a line. It also simplifies the discussion by
specifying one and the same background distribution for both fits.} (S=
123.4 counts, $\Gamma$ = 23.1 keV, $E_S$ = 792.8 keV), and has the
value $\chi^2$ = 16.65, a substantial reduction from the Background-Only
value of 28.30.  Because of the three additional parameters  of the
gaussian line, the $\chi^2$ distribution now has only 17 degrees of
freedom. The reduced chi-squared value, $\tilde{\chi}^2$, therefore
decreases to 0.98 from the 1.41 value for the Background-Only fit.
Clearly the One-Sharp-Line description is statistically superior to the
Background-Only description presented in the APEX report. The
statistical significance of that superiority is further quantified by
means of Confidence Level analysis for upper and lower bounds, as
discussed below. In addition, the $\pm$2.58$\sigma$ $\chi^2$ 99\%
interval, into which repeated measurements should fall with 99\%
probability, extends from about 24  counts to 252 counts. The
chi-squared analysis therefore implies it to be very unlikely that any
repetition of the APEX experiment could fail to produce excess sharp
pairs near 800 keV.

\widetext
\begin{center}
\renewcommand{\arraystretch}{1.2}
\begin{tabular}{|l|c|ccc|c|}  \hline 
\multicolumn{6}{|c|}{TABLE II: BEST $\chi^2$ FIT to APEX' U+Th PAIR DATA}\\ \hline \hline
& D$_f$=No.    &\multicolumn{2}{c}{Fit Quality:}   &Probability:  
&99\% Sh.Pair Range:\\
 &D.of F. &$\chi^2$,\hspace{2mm}  &$\tilde{\chi}^2=\chi^2/D_f$;\hspace{2mm}    &$P_{\chi}(\chi^2/D_f)$\hspace{2mm}    &\hspace{2mm}$N_{SP}^L\leq N_{SP}\leq N_{SP}^U$ \\ \hline
APEX' Background-Only: &20 &28.30 &1.41 &10.2\% &No Sharp Pairs    \\
Background+Sharp Line: &17 &16.65 &0.98 &47.8\% & $24\leq123\leq252$\\ \hline
\end{tabular}
\end{center}

\begin{flushleft}
\widetext
\footnotesize
Table II. The Table compares two fits to twenty one of the $(e^+e^-)\; $ pair
counts reported by the APEX collaboration in Ref.\cite{ahma/95}. The
three parameter Background+One-Sharp-Line fit is a better fit than
APEX'  one parameter Background-Only description, as discussed further
in the text. These results support the hypothesis that an excess of
sharp pairs occurs in the APEX data near 790 keV. Column six shows that
one should expect a re-measurement of the 790 keV excess sharp pair
count to yield a positive value in the range from 24 to
252 counts more than 99\% of the time.
\end{flushleft}

\normalsize

\narrowtext

When the assumed description is in fact the true description, then the
most probable reduced chi-squared value, $\tilde{\chi}^2$, is 1.0, and
larger and smaller vaues are expected to occur in subsequent
measurements with 50\% probability. Table II shows that for the
One-Sharp-Line distribution one expects a new measurement to yield a
larger chi-squared value 47.8\% of the time, as compared with the value
of 10.2\% for the Background-Only distribution.

\section{99\% Confidence Level Bounds on  APEX' Sharp  Pairs}

We next apply Confidence Level analysis to the APEX data of Table I and
Figure 2, to set upper and lower limits upon the mean excess sharp pair
counts. Specifically, we compute the bounds which the APEX data impose
(at the 99\% Confidence Level) upon the greatest mean value, $\nu_U$,
and upon the least mean value, $\nu_L$, of excess sharp pairs for 20-,
40-, and 60- keV intervals, corresponding to the treatment of the APEX
data bin by bin, and as sums of two and of three adjacent bins. The
resulting bounds for the twenty one 20 keV bins are isted in columns
ten and eleven of Table I.

The analysis assumes that the observed pair count in each bin  is
composed of the sum of a background contribution, $n_B$, arising
randomly from a Poisson distribution of mean value, $N_{BK}$,

\begin{equation}
P(N_{BK};n_B) = (exp-N_{BK})[(N_{BK})^{n_B})/n_B!],
\end{equation}

\noindent and a sharp pair contribution, $n_{SP}$, arising from a
second Poisson distribution of mean, $\nu$. The  mean background value,
$N_{BK}$, in each bin is taken from the APEX' best fitted event-mixed
distribution (given in column three of Table I), and the value of $\nu$ is  chosen to yield the
pre-selected Confidence Level (CL), that a repeat of the APEX
experiment would yield a count smaller than the observed count,
$N_{OBS}$. As a function of the mean sharp pair count, $\nu$, that
probability is defined\cite{hern/90,hele/83} as follows\footnote{The
present form of $\alpha$ is given by Eq(II.28) of Ref.\cite{hern/90},
which follows identically from Eqs.(2) and (7) of Ref.\cite{hele/83} by
integration and summation.}:

\begin{equation}
\alpha(\nu;N_{OBS})=[\sum_{n=0}^{N_{OBS}}(P(N_{BK}+\nu;n)]/[\sum_{n=0}^{N_{OBS}}(P(N_{BK};n)]
\end{equation}

As $\nu$ increases from zero to $\infty$, the value of $\alpha$
decreases from 1 to zero.  When  $\nu$ is such that the quantity
$\alpha(\nu;N_{OBS})$, has the value, 0.010, then that value is $\nu$ =
$\nu_U$, and is the 99.0\% Confidence Level upper bound upon the true
mean value, $\nu_{TRUE}$:  If $\nu_{TRUE}$ had the value $\nu_U$, then
99\% of repeated measurements would yield a smaller value for n than
$N_{OBS}$. Likewise when $\alpha(\nu)$ assumes the value 0.990, the
corresponding value of $\nu$ is equal to $\nu_L$ , the 99.0\% CL {\it
lower} bound upon the mean sharp pair count: If $\nu_{TRUE}$ had the
value $\nu_L$, then 99\% of repeated measurements would result in a
number, n, of counts greater than $N_{OBS}$.

Figure 3 plots the excess pair counts per 20 keV bin, $N_{EXC}$ (from
column five of Table I), defined as the difference, $N_{OBS} - N_{BK}$,
of columns three and two, together with the standard deviations, $\pm
\sigma_i = \pm N_{OBS}^{1/2}$, plotted as error bars. $N_{EXC}$
exhibits a maximum\footnote{For the best fitting One-Sharp-Line
distribution, this maximum in the one 20-keV bin count corresponds to a
total sharp pair line strength of 123.3 counts, as given in Table III.}
of 83$\pm$26.1 counts/20 keV for the bin at 790 keV, and is best
described (in the $\chi^2$ sense) by the gaussian one-line
distribution, shown cross hatched.  For this same 790 keV bin, the
figure also exhibits (as triangles) the 99\% CL bounds upon the mean
sharp pair value implied by its 83 count excess. The upper bound of the
790 keV bin count is 145.4 counts/20keV.

However, the greater interest for the present discussion lies in the
results for the {\it lower} bounds imposed by the data. For all of the
energy bins except that at 790 keV, the lower limits listed in Table I
have a  negligible value ($\leq$2 counts per 20 keV), indicating that
the APEX data fails to provide ``statistically significant'' (i.e., at
the 99\% CL) evidence to support even a positive mean excess pair count
even as small as two.  But for the 20 keV bin centered at 790 KeV, the
lower bound becomes substantial, with a value\footnote{For our best fit
distribution, this corresponds to a lower bound of 35.4 counts upon the
total line strength and an upper bound of 217.1, as given in Table
III.} of 24.1 counts per 20 keV (about 0.9 $\sigma$).

In summary, Figure 3 shows that for the measured 3.2$\sigma$ excess of
83 counts in the 20  keV bin at 790 keV, the Confidence Level analysis
indicates  a 99\% CL lower bound of 24 counts upon the sharp pair mean,
and a  99\% CL upper bound of 146 counts.  The APEX data therefore
provides statistically significant evidence that a non-zero positive
excess of sharp pairs above the background occurs in a 20 keV bin hear
800 keV.

Analogous 99\% upper and lower bounds have been established by
considering also the counts/40keV and  the counts/60 keV which arise by
combining two and three adjacent 20 keV bins.  The results are
presented in Table III. For each bin grouping a total line strength
also  has been inferred by assuming our  best fitting line shape.  Each
of the four analyses provides statistically significant evidence that
the mean total excess sharp pair count near 800 keV is greater than 23
counts and less than 252 counts.

\widetext
\begin{center}
\renewcommand{\arraystretch}{1.2}
\begin{tabular}{|l|c|c|c|}  \hline 
\multicolumn{4}{|c|}{TABLE III: 99\% CONFIDENCE LEVEL UPPER and LOWER BOUNDS}\\
\multicolumn{4}{|c|}{on APEX' SHARP  800 keV PAIR COUNTS}\\ \hline \hline
\multicolumn{1}{|c|}{Data} &Pair  &$\nu_L<(N_{EXC}\pm \sigma)<\nu _U$  &$\nu_L/\sigma<(N_{EXC}/\sigma)<\nu_U/\sigma$ \\
\multicolumn{1}{|c|}{Grouping} & Energy(keV) & (counts)&(Stnd. Deviations) \\ \hline \hline
Single Bin (20keV) & 790 & $24.1<(83.0\pm26.1)<145.4 $ & $0.9<3.2<5.6$\\ \hline
\multicolumn{4}{|c|}{$(f_1=0.673)\Rightarrow ($Total Line Strength$= 35.4<123.3<217.1$);}   \\ \hline \hline
Double Bin (40keV) & 800 & $29.4<(111.5\pm36.0)<197.8$ & $0.8<3.1<5.5$ \\ \hline
\multicolumn{4}{|c|}{$(f_2=0.901)\Rightarrow ($Total Line Strength$=   32.6<123.8<219.5$);} \\ \hline \hline 
Triple Bin (60 keV)&790 & $23.4<(123.4\pm43.7)<226.3$ & $0.5<2.8<5.2$ \\ \hline
\multicolumn{4}{|c|}{$(f_2=0.997)\Rightarrow($ Total Line Strength$= 23.4<123.6<226.7$);} \\ \hline \hline
$\chi^2$ Best Fit*: & 792.8 & $24.0<(123.4\pm45.8)<251.1$ &  $0.5<2.7<5.5$ \\\hline 
\multicolumn{4}{|c|}{*(Here a $\pm\sigma_{\chi^2}$ shift increases $\chi^2$ by +1, and the shift to the 99\% Interval, by +(2.58)$^2$.}\\ \hline
\end{tabular}
\end{center}

\widetext
\footnotesize
\begin{flushleft}
Table III presents the  (99\% Confidence Level) lower and upper bounds,
$\nu_L$ and $\nu_U$, which the APEX data imposes upon the mean excess
pair counts in the single 20 keV bin centered at 790 keV, in the
two-bin combination centered at 800 keV, and in  the three-bin
combination centered at 790 keV. For all of the 20 keV bins other than
that at 790 keV, the 99\% CL lower bound is negligible (less than two
counts), indicating that only near 790 keV does the APEX date {\it
require} a positive pair excess. For each grouping, the table also
presents a projected total pair count based upon the fraction f$_n$ of
excess pairs expected to occur in the interval in question if the
excess pairs are distributed according to the  gaussian One-Sharp-Line
$\chi^2$ best fit described in the text. The $\chi^2$ 99\%
($\pm$2.58$\sigma$) interval for the total line strength is also
presented. All of our analyses support a sharp pair total line strength
of about 123.5 counts, and statistically significant lower bounds equal
to or greater than 23 counts and upper bounds less than or equal to 252
counts.
\end{flushleft}
\normalsize

\narrowtext

We emphasize that the quantitative results of the present analysis
contradict in no way the quantitive empirical results of the APEX
report\cite{ahma/95}. In particular, their 99\%CL upper limit of 292
total sharp pair counts (0.572$\mu$b/sr in their Figure 2b) near 800
keV is consistent with, and in fact less restrictive\footnote{Neither
the APEX report\cite{ahma/95} nor M.  Wolanski's thesis\cite{wola/95},
where the upper bound analysis was first published, provide sufficient
detail to determine the source of the difference between their upper
bound and ours.} than any of the present upper bounds upon the total
counts\footnote{Table III shows that for the analyses of the 20-, 40-
and 60-keV bins, the upper bounds on the total sharp pair line strength
are 217, 220, and 227 respectively, and that the chi-squared 99\%
interval's upper limit is 252. We refer to them all together as
specifying the range (217 to 252) of upper bounds on the sharp pair
line strength.}. Moreover, the APEX upper bounds of Ref.\cite{ahma/95}
exhibit a maximum near 800 keV, like that which Table I presents for
the present upper bounds, which arises from  the large excess of pair
counts measured in the 790 keV bin.  Presumably the APEX analysis would
also have produced a positive 99\% CL {\it lower} bound, also greater
than that presented here, if one had been extracted.

\section{Discussion of APEX Positive Sharp Pair  Evidence} 

APEX reported its data and concluded that ``.....the results of the
present  experiment represent a real disagreement with the previous
observations''.  Yet the present analysis of the same data shows that
the APEX data corroborates the existence of sharp pairs at a confidence
level exceeding 99\%.  How can the same data support both conclusions?

Our answer to this question divides into two parts. In the first we
compare APEX' actual sharp and background pair counts against EPOS' in
terms of  strictly empirical data-only ratios of ``Qualitatively
Similar'' quantitities. We conclude that these pair ratios indicate not
only no contradiction between APEX and EPOS, but in fact evidence a
remarkable comparability and consistency between the EPOS and the APEX
pair databases, both in size and quality. In the second we exhibit
APEX' expectations for the sharp pair count as inconsistent with the
EPOS data whence it derives, and as excessively large. These
deficiencies are traced to the APEX assumption that the sharp pair
production cross section is constant and has the value of
5.0$\mu$b/sr.

\subsection{Purely Empirical Comparisons of APEX/EPOS Pair Databases}\ 

Given that the measured APEX data supports a sharp pair line near 800
keV, one must ask whether its observed 123 count strength is consistent
with the EPOS' sharp pair count (97$\pm38$ counts, FWHM= 40 keV) at 809
keV.  We here present a purely empirical positive answer in terms of
data-only ratios of ``Qualitatively Similar'' pair counts. These ratios
indicate that for ``Qualitatively Similar'' pair counts  the APEX and
the EPOS U+Th pair data bases are quite comparable both in size and
quality, and they suggest no contradictions between them. \

\subsubsection{The Requirement of Qualitative Similarity}

We first digress to emphasize that it in any quantitative comparison of
the APEX and EPOS pair experiments which claims to be purely empirical,
the APEX pairs must be of the type accepted in the EPOS experiment.  We
refer to this as the requirement of ``Qualitative Similarity''. Without
Qualitative Similarity, the expression for a ratio of comparable
measured pair counts from the two experiments involves a ratio of the
underlying cross sections for the pair production processes which is
not a measured quantity. But for ``Qualitatively Similar''
processes the underlying unknown cross sections are one and the same,
and their ratio is known to be equal to one, despite the fact that they
are unmeasured. Thus, a violation of  the requirement of ``Qualitative
Similarity'' introduces an unmeasured ratio of two different production
cross sections. If this ratio is unknown, then it defeats the
calculation of the ratio\footnote{If the pair production process were
well understood, theoretical information could be supplied to fix the
ratio. The comparison might then be credible but could not be
considered purely empirical in the strict sense of the term.}. If an
additional non-empirical assumption is made about the ratio, then it
becomes ambiguous whether the comparison should be viewed as addressing
the original question or the additional assumption.

For example, one cannot know empirically whether the APEX'
comparison\cite{ahma/95} of their own (1,n) pair set (in which any
number n of electrons are accepted in coincidence with one positron)
with EPOS' (1,1)-only pair set (in which  pairs only from events in
which one and only one electron is in coincidence with the positron are
accepted) ought to provide evidence on the existence of sharp pairs or simply
on the differences between sharp pair production from (1,1) events and
from from (1,n) events. If one {\it assumes}, implicitly or explicitly,
that the probability of sharp pairs is the same in the (1,n) events as
it is in the (1,1) events, then a result can be obtained, but its
reliability is contingent upon the correctness of the assumption, and
the analysis can no longer be considered as purely empirical.

We therefore consider in Table IV only APEX' counts of pairs which are
of the ``EPOS-type'': the leptons must be observed  in opposite arms of
the experiment (``RL'' pairs), and the pairs must arise only from
events in which one positron and one and only one electron were
emitted, (``(1,1) pairs'')\footnote{Wedge cut pairs are not considered
in this comparison because the APEX' wedge differed somewhat from
EPOS'.}.

\subsubsection{EPOS', APEX' Qualitatively Similar Pair Databases Are Nearly  Equivalent}

Table IV summarizes some quantitative characteristics of the APEX and
EPOS U+Th experiments, including their total luminosities and selected
characteristics of their Qualitatively Similar RL(1,1) pair sets:
their pair count totals and their pair counts per 20 keV near 800
keV.  Also presented in the last two columns are the observed counts,
$N^{800}_{SP}$, of excess sharp pairs and the ratio of these to background counts
per 20 keV near 800 keV.  

Table IV shows that despite the APEX' $\sim3\times$ larger luminosity, it
collects only about the same number of coincident EPOS-type pairs as
EPOS, both overall and in  20 keV intervals near 800  keV. These ratios
indicate that  the overall size and the general shape of the APEX
background  pair distribution is similar to that of EPOS. It follows
that in direct comparisons of APEX' and EPOS' measured pair data, APEX
can claim no clear statistical superiority over EPOS.

Besides the counts of their background distributions, the  numbers of
excess sharp pairs near 800 keV and their ratios to the background pair
counts near 800 keV in columns five and six are also comparable for the
two experiments. Thus Table IV shows that as regards both their
coincident pair background distributions and their excess sharp pair
distributions\footnote{We note that the APEX sharp pairs have been
extracted in the present analysis from the published APEX data, which
includes (1,n) pairs not of the EPOS type.  The ratios in columns five
and six of Table IV, line 3, therefore violate the requirement of
``Qualitative Similarity''. Since the selection of  the (1,1) subset
will surely reduce these pair counts,  they can nevertheless provide
upper bounds upon the corresponding Qualitatively Similar ratios.
Clearly this ratio should be replaced by the analogous ratio for the
excess RL(1,1) pairs above the RL(1,1) pair background when that data
becomes available.}, the APEX and EPOS data bases are comparable within
$\sim$30\%;  Table IV therefore not only speaks against any substantial
contradiction between them, but it supports their overall mutual
consistency. It also precludes any claim of statistical superiority for
the APEX' pair data over EPOS', and raises the question why  APEX
counted so few pairs (both of the smooth background and sharp 800 keV
types) compared with EPOS despite its larger luminosity, a question to
be discussed further below.

\widetext
\begin{center}
\renewcommand{\arraystretch}{1.2}
\begin{tabular}{|l|c|cc|cc|}  \hline 
\multicolumn{6}{|c|}{TABLE IV: COMPARATIVE MEASURES OF  APEX and EPOS PAIR DATA BASES}\\ \hline \hline
&L$^{TOT}$($\mu$b$^{-1}$) &$N_{TOT}^{LR(1,1)}$   &$\Delta N_{800}^{LR(1,1)}/20keV$   &$N_{SP}^{800};$ &$(N_{SP}^{800})/(\Delta N_{800}^{LR(1,1)}/20keV)$  \\ \hline
APEX: &7000$^{(a)}$ & (40.8K)$^{(c)}$  &(1480)$^{(c)}$  &$\leq$123$\pm$36$^{(d)}$ &$\leq$0.083 \\
EPOS: &2196$^{(b)}$  & 50K    &1280  &97$\pm$38 &0.076 \\ \hline
(APX/EPS): &3.2   &  0.8     &1.2    &$\leq$1.3 &$\leq$1.09\\ \hline
\end{tabular}
\end{center}

\widetext
\footnotesize
\begin{flushleft}
Table IV. The APEX/EPOS total luminosity (L$^{TOT}$) ratio inferred
from the positron yields and efficiencies is substantially
($\approx3\times$) larger than the corresponding ratios ($\approx1$) of
comparable pairs counted overall and per 20 keV near 800 keV. However, for
the RL(1,1) pairs of the type which EPOS accepted, the APEX pair data
base is somewhat smaller overall and somewhat larger near 800 keV, but
not 3X larger as its luminosity would suggest. Furthermore, the ratio of
the observed APEX and EPOS {\it sharp} pair counts near 800 keV is
quite consistent with the ratio of background pairs in the same energy
range. This table offers no evidence of any contradiction between APEX'
and  EPOS' sharp pair results. To the contrary it presents two very
similar pair distributions which quite consistently provide two very
similar sharp pair signals near 800  keV. The Table does raise the
question why APEX counted so few EPOS-type RL(1,1) pairs as compared
with EPOS. Notes to the table entries follow:  (a) APEX' measured
luminosity\cite{ahma/95}; (b) EPOS' luminosity inferred in Table V by
comparison of its positron data with APEX', following the method of
Cowan and Greenberg\cite{cowa/96}; (c)  To guarantee Qualitatively
Similar APEX/EPOS comparisons (see text) we consider only APEX
pairs of EPOS' RL(1,1) type.  These RL(1,1) data are taken from
Ref.\cite{wola/95}, Figures VI.2.2, and VI.2.2(c); see also Note 18; (d) See Note No.19.
\end{flushleft}

\normalsize

\narrowtext
\subsection{APEX' Expectations are Inconsistent, and Excessive}

But if the APEX and EPOS pair databases are essentially  equivalent, how
does one understand the APEX' claim that its data contradicts the EPOS
data?  The answer is that APEX' contradiction is in fact not with the
EPOS data but with APEX' own expectations,  which are in fact based not
upon the EPOS data, but upon a  misconstrual of that data. We therefore
consider here what APEX analysis did to generate its expected pair
counts, what it might have done, and what the effects were. The
conclusion is that APEX, purely by an inconsistent assumption,
multiplied its sharp pair expectations by an order of magnitude over
those which the EPOS data can actually sustain.

To set its expectations for its own results, APEX chose to epitomize
the EPOS 809 keV line by the ``maximal'' value of the sharp pair
production cross section reported\footnote{This value was presented by
EPOS\cite{sala/90}, without further specification, as a rough generic
value for all of its lines. More recently the EPOS/II
collaboration\cite{ganz/96} has published (also without specification
of the averaging interval) the value of 1.4$\mu$b/sr, which would seem
to be appropriate for the 0.07 MeV/U single run beam energy spread of
the original EPOS experiment.} as ``on the order of 5.0$\mu$b/sr''.
But APEX chose to disregard EPOS' evidence\footnote{EPOS observed sharp
pairs near 800 keV only in two of its five runs at different energies.
One straightforward interpretation of those results would be that the
sharp pair production process is beam energy dependent, and occurs {\it
only} in a narrowly resonant energy interval which falls into the 0.02
MeV/U , $\approx$4.8 MeV, range from 5.85 MeV/U to 5.87 MeV/U. The
implications of this as an alternative assumption to the APEX' are
exhibited in Table V.B.} of a  beam energy dependence of the sharp pair
production, and assumed that the cross section is also

\widetext
\begin{center}
\renewcommand{\arraystretch}{1.2}
\begin{tabular}{|l|c|crclccc|c|}  \hline 
\multicolumn{10}{|c|}{TABLE V: EXPECTED SHARP PAIR COUNTS}\\ \hline \hline
\multicolumn{10}{|l|}{A: CONSTANT 5.0$\mu$b/sr CROSS SECTION:}\\ 
\multicolumn{10}{|c|}{$N_{SP}= L^{TOT}\overline{<d\sigma_{SP}/d\Omega_{HI}>}\Delta\Omega_{2I}^{eff}G_{SP}$}\\ \hline
 &$N^{OBS}_{SP}$ &L$^{TOT}$ & $\times$ &$\overline{<d\sigma/d\Omega>}$ & $\times$ & $\Delta\Omega_{2I}^{eff}G_{SP}$ &$=$ &N$^{EXP}_{SP}$ &N$^{EXP}_{SP}/N_{OBS}$ \\ \hline
A1.APEX:    &123  & 7000 &&5.0   &&0.0704   &&\{2464\} &20.0 \\ \hline
A2.EPOS:    &97 &2196  &&5.0   &&0.0855  &&\{939\}  &9.7 \\ \hline \hline
\multicolumn{10}{|l|}{B: CROSS SECTION NON-ZERO ONLY NEAR E$_R$ = 5.86 Mev/U:}\\ 
\multicolumn{10}{|c|}{$N_{SP}=\tilde{L}(E_R)\Delta\Omega_{2I}^{eff}G_{SP}\int_{\Delta E_R}<d\sigma(E)/d\Omega>dE$}\\ \hline 
&$N^{OBS}_{SP}$ &$\tilde{L}(E_R)$ &$\times$&$\int<d\sigma(E)/d\Omega>dE$ &$\times$& $\Delta\Omega_{2I}^{eff}G_{SP}$ &$=$ &N$^{EXP}_{SP}$ &N$^{EXP}_{SP}/N_{OBS}$ \\ \hline
B1.EPOS: &97 &12.5K &&\{0.091\}  &&0.0855  &&97  &1.0 \\ \hline
B2.APEX: &123 &41.2K   &&0.091 &&0.0704  &&\{264\} &2.1\\ \hline \hline
\multicolumn{10}{|l|}{C: EPOS' LUMINOSITY from POSITRONS and APEX' LUMINOSITY:}\\ 
\multicolumn{10}{|c|}{$N_{e^+}= \L^{TOT}\Delta\Omega_{2I}^{eff}G_{e^+}\overline{<d\sigma_{e^+}/d\Omega_{HI}>}$}\\ \hline 
&$N^{OBS}_{e^+}$ &$L^{TOT}$ &$\times$&$\overline{<d\sigma_{e^+}/d\Omega_{HI}>}$ &$\times$& $\Delta\Omega_{2I}^{eff}G_{e^+}$ &$=$ &N$_{e^+}$ &$N^{OBS}_{e^+}$/$N_{e^+}$ \\ \hline
C1.APEX: &246K &7000 &&\{173\}  &&0.2031    &$=$&246K &1.0 \\ \hline
C2.EPOS &250K  &\{2196\} &&173  &&0.6580    &$=$&250K &1.0\\ \hline \hline
\multicolumn{10}{|l|}{D: APEX and EPOS PARAMETER VALUES for SHARP PAIRS and POSITRONS:}\\ \hline
&$\Delta\Omega_{2I}^{eff}$ &$\epsilon_{X}$   &$\times$& LT   &$\times$& $W_{X}$ &$=$ &$G_{X}$ &$\Delta\Omega_{2I}^{eff}G_{X}$  \\ \hline
1)APEX(SP):    &6.86  &1.3\%  &&0.8   &&0.987  &&0.0103    &0.0704 \\ \hline
2)EPOS(SP):    &7.03 &1.4\%  &&0.9   &&0.965  &&0.0122    &0.0855 \\ \hline 
3)APEX(e$^+$): &6.86  &3.7\%  &&0.8   &&1.00   &&0.0296    &0.2031 \\\hline
4)EPOS(e$^+$): &7.03  &10.4\% &&0.9   &&1.00   &&0.0936    &0.6580 \\\hline
\end{tabular}
\end{center}

\widetext \footnotesize \widetext Table V (in part A) displays the
numbers of sharp pair counts expected in the the APEX and the EPOS
experiments under the APEX' constant 5.0$\mu$b/sr cross section
assumption, and (in part B) under the alternative assumption that the
pair production cross section is non-zero only in a narrow beam energy
interval lying between 5.85 and 5.87(MeV/U).  Part (V.C) presents the
calculation of EPOS' luminosity, and Part (V.D) summarizes the various
parameter values used in the calculations.  In each line the quantity
extracted is placed in brackets \{\}.  The Table shows that APEX'
constant 5.0$\mu$b/sr assumption produces an expected pair count (2464:
line A1) for APEX which is an order of magnitude larger than the
alternative assumption (264: line B2), and, correspondingly for EPOS a
count, (939, line B2), which is an order of magnitude larger than the
pair count which EPOS actually observed (97: line B1). The cause of
these large expected values is discussed in the text. In part B, the
table shows that when the energy integral of the pair cross section is
fixed by the EPOS sharp pair count and by the EPOS luminosity density
from the EPOS/APEX positron counts of part C, the APEX expectation is
reduced by $\approx10\times$ to 264, from the 2560 which APEX expected.
(See Notes 1 and 8.) This value is somewhat smaller than the upper
bound (292 counts) set by the APEX' experiment, and about 2.1 times
larger than the sharp pair count which APEX actually observed.  Part C
of the table shows how the average positron cross section is obtained
(line C1) from the APEX positron data, and (line C2) how the EPOS
luminosity, $L^{TOT}$, is inferred from that cross section and EPOS'
positron data. Part D of the table collects the values of the various
experimental parameters used in these calculations:  the effective
two-ion solid angle, ($\Delta\Omega_{2I}^{eff}$), the  detection
efficiencies, ($\epsilon_{X}$), for sharp pairs and for positrons,  the
``live time''factors, (LT), the wedge factors (See Note 22.),
($W_{X}$), and the gathering powers, $G_{X}$, defined in the table.
$G_{X}$ gives the fraction of the emissions of type X created at the
target which are actually counted by the detector. The units are
$\mu$b, sr, MeV/U and combinations of them.\ \vspace*{0.1in}
\renewcommand{\baselinestretch}{1}
 \normalsize

\narrowtext

\noindent  constant in energy. Such a constant 5.0$\mu$b/sr cross
section assumption, together with the APEX' luminosity of
7000$\mu$b$^{-1}$ predicts\footnote{(using appropriate parameters as
given in Table V.D to characterize the apparati.) Note also that the sharp
pair wedge factors in Table V.D are computed on the assumption that the
sharp pair distribution is a gaussian in the difference energy with the
measured width\cite{sala/90,grif/91b}.} (in Line A1, Table V) that the
APEX experiment should count $\sim$2464 excess sharp pair
counts\footnote{APEX' published simulation provided a gaussian best fit
of 2560 near 800 keV. Our value here of 2464 is equivalent to that
result for the purposes of the present discussion. See also Notes 1 and
7.}.

\narrowtext
However, it is easy to see that this estimation is physically
inconsistent. One simply applies it to compute the EPOS sharp pair
count,  whence the assumption derived.  Using EPOS'
luminosity\footnote{This is the value calculated in Table V.C from APEX
luminosity and the EPOS and APEX positron counts.} of
2196$\mu$b$^{-1}$, this calculation requires (in Line A2, Table V) that
EPOS should count 939 excess sharp pairs, whereas in fact\footnote{This
inconsistency was first pointed out  in Refs.\cite{grif/96se}, and is
also noted in Ref.\cite{cowa/96}.} EPOS counted only 97. Thus APEX'
constant 5.0$\mu$b/sr assumption immediately contradicts the EPOS data
which it was supposed to epitomize. To make the prediction for EPOS'
pairs consistent with the actual EPOS count, the APEX' constant cross
section\footnote{The APEX' reply\cite{ahma/96} to Cowan and
Greenberg\cite{cowa/96} expresses the opinion that ``a proper
discussion...should be carried out using apparatus independent cross
sections''.  But cross sections are not apparatus independent when the
beam energy spread is larger than the range over which the cross
section is constant. In the limit when the cross section is non-zero
only in a narrow beam energy interval within the beam energy spread, it
is the energy integrated cross section, not the cross section itself,
which becomes ``apparatus independent''. The useful ``cross section''
in this case is an effective average cross section defined to make the
energy integral of the cross section over the beam spread of any given
apparatus equal to the correct physical energy-integrated cross
section. This cross section is obviously apparatus dependent, since it
must vary inversely with the beam energy spread. In particular, a given
thin target EPOS ``cross section'' must then be assigned a different
effective value when it is measured with APEX' thick target beam. E.g.,
the energy integrated value of EPOS' pair cross section required (See
Table V.B) to give its $\sim$100 counts is roughly
0.1($\mu$b/sr)(MeV/U).  Thus, EPOS' \cite{sala/90} reported ``cross
section'' of 5$\mu$b/sr would correspond to a 0.02 MeV/U beam energy
interval  (such as the interval from 5.85 to 5.87 MeV/U, where a narrow
production cross section would contribute to the two EPOS runs where
sharp pairs were observed). The more recently stated (in Table I of
Ref.\cite{ganz/96}) value of 1.4$\mu$b/sr is an appropriate average for
the EPOS' full single run beam spread of 0.07 MeV/U.  The values of
0.62 and 0.59$\mu$b/sr, for the full 0.16 and 0.17 MeV/U spreads of the
EPOS and APEX experiments respectively,  are also appropriate
characterizations of the same EPOS 809 keV sharp pair count. Under
these circumstances, the entire APEX discussion, which overlooks all of
these possibilities, becomes ambiguous and inconsistent.} would have to
be reduced from 5.0$\mu$b/sr to 0.52$\mu$b/sr, a value somewhat smaller
than  the APEX' 99\% CL upper bound of 0.572$\mu$b/sr.

If, instead of assuming a constant cross section, APEX assumed what the
EPOS data suggests, that the cross section is non-zero only in  a
narrow energy range within the 0.02(MeV/U) beam energy interval from
5.85 to 5.87 (MeV/U), then its expectations would be altered
substantially. In this case EPOS' luminosity density\footnote{This
value is equal to 5.7\% (per 0.01 MeV/U) of the total EPOS luminosity.
Because EPOS' five runs overlapped more at some beam energies than at
others, its luminosity density varies with beam energy, as illustrated
in Fig. 4. A simplified representation of EPOS' luminosity density was
given in Fig. 1a of Ref.\cite{ahma/96}, which would yield an EPOS
luminosity density 9\% too large at 5.86 MeV/U.}  near 5.86 MeV/U,
[12.5K($\mu$b$^{-1}$)(MeV/U)$^{-1}$], and its 97 sharp pair counts
imply (in Line B1, Table V) an energy-integrated  sharp  pair cross
section of 0.091($\mu$b/sr)(MeV/U) as the apparatus independent
characteristic of the EPOS sharp pairs. From this integrated cross
section, APEX' luminosity density,
[41.2K($\mu$b$^{-1}$)(MeV/U)$^{-1}$], implies (in Line B2, Table V)
that 264, not 2464, sharp pairs should be seen near 800 keV.\

We note that in this case the average cross section from 5.85 to 5.87
(MeV/U) is 4.6$\mu$b/sr, a value ``of the order of'' the 5.0$\mu$b/sr
``maximal'' cross section mentioned by EPOS\cite{sala/90}. The APEX'
extrapolation of the 5.0$\mu$b/sr  cross section to the whole of its
0.17(MeV/U) beam energy interval then assumes an energy integrated
cross section of 0.85($\mu$b/sr)(MeV/U), instead of the actual value of
0.091, and effects thereby an unjustified  multiplication of the
energy-integrated cross section by a factor of (5.0/4.6)(0.17/0.02) =
9.3. Their sharp pair expectations are multiplied also by this same
factor. It  would appear that this extrapolation is the cause of APEX'
expectation (line A1, Table V) that its 40.8K RL(1,1) pairs should
exhibit $\sim$2500 sharp pairs near 800 keV, while EPOS' 50K pairs
exhibited (line B1, Table V) only 97, and of its physically
inconsistent and excessively large expectation (line A2, Table V) for
the EPOS observed pair count. \

Thus, by assuming a constant cross section, despite EPOS'evidence for
an energy dependence, and by assigning it an average value of
5.0$\mu$b/sr appropriate only for a narrow beam-energy interval, the
APEX analysis multiplies its expected pair count by an order of
magnitude\footnote{This assumption also leads APEX to
estimate\cite{trai/96} the EPOS luminosity to be $\approx$400
$\mu$b$^{-1}$, some 5.5 times smaller than the value of 2196
$\mu$b$^{-1}$ obtained from the independent positron data in Table V.
Thus, besides inflating the expectations, APEX assumption encourages
the misapprehension that APEX luminosity is some 17.5 times larger than
EPOS, rather than 3.2 times larger.}. \

\subsection{APEX' Apparent Detection Efficiency for Backround Pairs is Much
Smaller than EPOS'}

An analysis similar to that of Table V can be used to obtain a purely
empirical estimate of the ratio of EPOS' to APEX' overall detection
efficiencies for smooth background pairs. Consider the values of
$N_{TOT}^{LR(1,1)}$ given in Table IV, the relationship,

\begin{equation} 
N_{TOT}^{LR(1,1)} =  L^{TOT}\Delta\Omega_{2I}^{eff}(LT)\epsilon_{BP}\overline{<d\sigma_{RL(1,1)}/d\Omega_{HI}>},
\end{equation}
 
\noindent and the values of $L^{TOT}$, $\Delta\Omega_{2I}^{eff}$, and
LT given in Table V.D. Then under the assumption that the ratio of the
energy and angular averages,
$\overline{<d\sigma_{RL(1,1)}/d\Omega_{HI}>}$,  of the cross section
for producing RL(1,1) background pairs is $\approx1$ in the two
experiments\footnote{At this point an attempt to compare an APEX (1,n)
pair count with a  Qualitatively Dissimilar EPOS (1,1) pair count would
fail because the ratio of the corresponding cross sections is unknown,
and  certainly not $\approx1$. In the present case the cross
sections are identical and one can reasonably assume that the that the
two averages differ from one another only slightly as a result of
specific detailed differences between the two generally similar
experiments.}, one computes the value of the ratio,

\begin{equation}
\epsilon^{EPOS}_{RL(1,1)BP}/\epsilon^{APEX}_{RL(1,1)BP}\;\approx\;3.38 \\ 
\end{equation}

\noindent and comes to the unexpected conclusion that the apparent value of EPOS' overall efficiency
for RL(1,1) smooth background pairs is more than $3\times$ larger than APEX'.

\subsection{APEX' Sharp Pair Count Imposes Small Upper Bound
Upon Its Sharp Pair Efficiency}

Of the several parameters assembled in Table V.D., the sharp pair efficiencies 
are least directly supported by laboratory measurement. Therefore, if a 
measured count of APEX' sharp  RL(1,1) pairs near 800 keV were available, 
a purely empirical estimate for the  ratio of EPOS' to APEX' sharp pair
detection efficiency would be of special interest.  However, the APEX'
sharp pair count of  Table IV includes pairs from (1,n) events of all n, and 
can therefore provide only an upper bound upon the RL(1,1) sharp pair count:
$N^{APEX}_{RL(1,1)SP}\:\leq\:N^{APEX}_{RL(1,n)SP} = 123\pm46$. Then the calculation provides the following bound upon the apparent ratio,

\begin{equation}
\epsilon^{EPOS}_{RL(1,1)SP}/\epsilon^{APEX}_{RL(1,1)SP}\; \geq \;2.2.
 \end{equation}

It is interesting that this result is in direct contradiction with the ratio
1.4/1.3= 1.07 of the published APEX\cite{ahma/95} and EPOS\cite{cowa/96}
sharp pair efficiencies (tabulated in lines D1 and D2 of Table V.D). It 
suggests that the inconsistency between APEX' observed  pair count of 123,
and the value of 264 expected by the calculation of Table V may arise
from the use of an erroneous value of a sharp pair detection efficiency, $\epsilon_{RL(1,1)SP}$, in the calculation
of the expected value.

\subsection{Alternative: Front Layer Production of (1,1) Pairs}

But an alternative possibility is that the ratios in equations (6) and
(7), although truly an evidence of inconsistency between the EPOS and
APEX experiments under the interpretation of equation (5), do not speak
so directly to discrepant detection efficiencies, but instead may be
indicating that the assumptions underlying equation (5) are
inappropriate for the RL(1,1) (both background and sharp) pairs.

Specifically, evidence has been published\cite{stie/87} that the most
probable ionization charge of a deflected\footnote{In Ref.\cite{stie/87}, Steibing et al. studied
Pb+Au, at deflections of 35$^\circ$ in the lab with target thicknesses
from 20$\mu g/cm^2$ to 870$\mu g/cm^2$.} high-Z projectile is
substantially larger in the layer of the target near its front face
than the mean equilibrium value which obtains after the projectile
penetrates deeper into the target\footnote{The author is grateful to
Dr. T.E.Cowan for pointing out these possibilities and the results of Stiebing et al.}. In this context one
could consider the hypothesis\footnote{More discussion is offered in Ref.\cite{grif/95i}.} that the (1,1) pair events are sensitive
to the projectile's ionization  charge, and more likely to occur in the
front $\sim$0.02 MeV/U of target.  In this case effective luminosities
would have to be used in equation (5), which are reduced by factors of
$\sim$0.02/0.17, and $\sim$0.02/0.07 for APEX' 0.17 MeV/U, and EPOS' 0.07 MeV/U
thick targets, respectively. Then the ratios of apparent pair detection
efficiencies in equations (6) and (7) would both be reduced by a
factor, 0.17/0.07$\sim$2.4, to the more tolerable values of 1.4 and
$\geq$0.9, respectively.

For this reason we have referred above to the efficiencies in equations
(6)and (7) as ``apparent'' efficiencies. An additional preference by
the (1,1) pair creation process for a particular narrowly resonant beam
energy interval, if it occurs in combination with such a ``Front
Layer'' preference, would require even further adjustments to the
proper effective luminosities to be used when equation (5) is
describing (1,1) pair events.

\section{A Current Perspective on the Sharp Lepton Problem }

\subsection{Substantial Experimental History }

\subsubsection{Sharp Positrons}
 
Regarding the larger inferences which ought now to be drawn, we
emphasize that a great variety of published data exists, not just on
coincident sharp pairs, but on their forerunners,  the sharp positrons
from high-Z collisions. In 1983, the first report\cite{schw/83} of a
sharp (FWHM$\sim80 keV$) positron line was published by the EPOS
collaboration. In 1985, evidence from six $Z_U=Z_1+Z_2$ combinations
showed persuasively\cite{cowa/85} that the energies of these positrons
were nearly independent of $Z_U$, and could therefore not be the
sought-after positrons expected \cite{rein/81} from the spontaneous
pair decay of the vacuum at $Z_U>Z_{crit}\approx172$ (whose energies
would increase rapidly with $Z_U$). By 1987, twenty three such positron
lines had been measured and tabulated\cite{koen/87}. Their energies
separated into three main groups, with mean energies of 255keV, 337KeV
and 396KeV.

\subsubsection{Sharp Pairs} By 1986, the EPOS' solenoidal spectrometer
had become a pair spectrometer and reported\cite{cowa/86} a narrow
coincident $(e^+e^-)\;$ pair line with summed energies near 760 keV, of
which the corrresponding positron projection resembled a narrow
positron singles distribution previously measured with the EPOS
apparatus.  In 1988, the Orange group, using its new double-orange
apparatus, reported\cite{berd/88} evidence ($4\sigma$) for a narrow
$(e^+e^-)\; $ line at a summed energy 815 keV and with opening angles
$\theta_{+-}\approx180^\circ$. The positron portion of this spectrum
also resembled a positron line they had measured previously. Ten
additional narrow $(e^+e^-)\; $ pair lines were reported by
Orange\cite{koen/89} in 1989, five others by EPOS\cite{sala/90} in
1990, and another seven lines by ORANGE\cite{koen/93} in 1993.  The six
EPOS lines group into three sets, near 625, 750 and 810 keV. The
seventeen Orange lines group similarly, but the energies are more
scattered and additional groups occur near 560 keV (4 lines) and 895
keV (2 lines).  \

\subsubsection{Sharp Leptons from Positron Bombardments} Furthermore,
Sakai \cite{saka/88,saka/91,saka/93}, has repeatedly reported {\it very
very} sharp (FWHM$\leq$ 3keV) 330 keV electrons from thin target
$[\beta^+ + U(or\; Th)]$ experiments following the thick target
experiments of Erb, et al\cite{erb/86}, and Bargholz, et
al\cite{barg/89e}, which first evidenced such pairs\footnote{Other
experiments\cite{wang/87,peck/87} did not find sharp pairs, but Sakai's
are the only thin target experiments, and the only experiments which
were repeated.  Also  Sakai argues\cite{saka/92a} that because of
differences in the experiments these null signals do not constitute a
contradiction of his results.}.

\subsection{Special Interest in Data from Alternatives Non-Heavy Ion Processes}

\subsubsection{Few MeV e$^+$/e$^-$ on High-Z Atoms}

Studies like Sakai's  warrant particular interest, since they do not
require a high-Z heavy ion collision and could be carried out
independently in many low energy nuclear and atomic laboratories at
comparatively modest expense. We have therefore analysed Sakai's data
elsewhere\cite{grif/95e,grif/96se,grif/95i} to infer a cross section
$\sim10^2$mb to produce the composite particle source of its sharp
electrons from a beam few MeV positrons on U (or Th). Furthermore, the
brehmsstrahlung part of this process should be independent of the charge sign
of the impinging particle, so that few MeV electrons could be as
well used as positrons to check Sakai's $\beta^+$ results.

This situation strongly recommends experiments with beams of
positrons and (equally well) of electrons in the energy range from 2 to
4 Mev, which can reliably measure a cross section of $10^2$ mb to yield
separately sharp ($\Gamma\leq\;2.1\;keV$) 330.1 keV electrons and/or
(their expected partner) positrons. Such a project could verify Sakai's
sharp line, and guarantee that it arises, at least in part, from the
brehmsstrahlung creation process\footnote{A supplementary experiment,
with positrons (only) of kinetic energy in the range, $660\leq\;
K^+\leq\;795\;keV$, could measure whether the composite particle pair
source can also be created by the Recoilless Resonant Positron
Absorption process first proposed in Ref\cite{grif/94} . Here, however,
Sakai's implied cross sections\cite{grif/95e} vary more widely and so
offer less reliable estimations of the magnitude of the cross section.}.

In addition, if Sakai's lines are ultimately to be encompassed
\cite{grif/94} by the Quadronium Composite Particle
Scenario\cite{grif/91a,grif/91b,grif/92a} for the Sharp Lepton Problem,
then the same scenario suggests\cite{grif/95e,grif/96se,grif/95i} that photons
resonantly absorbed on Uranium might produce $Q_\circ$ decay pairs
analogous to those which provide Sakai' sharp electrons, as we now
discuss.

\subsubsection{Pair Branching from Delbr$\ddot{u}$ck Uranium Resonances?}

In particular,  measurements of the {\it branching ratios}
(specifically, to pairs vs. gamma re-emission) for the sharp resonances
recently reported by Zilges, et al\cite{zilg/95} in U($\gamma\gamma$)
Delbruck scattering in the neighborhood of 1.8 MeV, promise to provide
pivotal information on the structure of these resonant states.  The
question is whether all of the resonance states are conventional
nuclear excited states, or whether one or more of them is a bound
supercomposite \{U,Q$_\circ$\} molecule in which the Q$_\circ$
composite particle source of the EPOS pairs is bound to the U atom, as
hypothesized\cite{grif/91a,grif/91b} in the ``Composite Particle
(Q$_o$?) Scenario'' for the EPOS data\footnote{This hypothesis allows
the decays from such molecular bound states to exhibit positive lepton
difference energies such as the EPOS U+Ta data require, and provided
the first positive suggestion that the decaying particle must be charge
composite.}$^,$\footnote{The experimental exclusion of all possible
lifetimes for such a composite particle asserted in Ref\cite{Tser/91}
emerged under analysis\cite{grif/93a,grif/93b} to be not an
experimental result at all, but a direct consequence of the unsupported
assumption of one channel decay (to (e$^+$e$^-$) {\it only})
  made in the analysis.}.

Experimentally, these alternative states should be signatured
qualitatively by their decay patterns: Q$_o$ ought to decay
preferentially to pairs, whereas nuclear dipole states decay primarily
by photon emission. Therefore, if one or more of these resonances is
measured to have a larger than expected pair emission
branch\cite{grif/96se,grif/95i}, then further investigation would be
indicated to determine whether the resonant state might be a
supercomposite \{U,Q$_o$\} state.

A parallel effort could study other U isotopes to determine whether (as
the Q$_o$ scenario suggests for the \{U,Q$_\circ$\} molecule), any
observed resonance is found also in the other isotopes of the same element:
$^{236}$U is {\it not} expected to exhibit the same {\it nuclear}
excitations as $^{238}$U, but it may support  the same \{U,Q$_\circ$\}
{\it supercomposite molecular bound state}.

Experimental studies of pairs from these processes therefore promise to
corroborate and extend (or to contradict and delimit) the presently
very limited  non-heavy-ion data evidencing sharp $e^+e^-$ pairs.
Whether in the end such experiments support or negate Sakai's sharp
lines, they will do so from an independent experimental platform and
they will be subject to independent verification in many laboratories.
In that respect they offer an invaluable tool for escaping the impasse
which seems to be developing among the heavy ion data.

\subsection{Quadronium Scenario, and Quantum Electrodynamics}

Our discussion has  been guided by the Composite Particle Scenario,
because it is the only hypothesis  which seems able to encompass the
whole range of data of the Sharp Lepton Problem. In that context, since
no experimental evidence presently speaks to the specific structure of
the composite particle,  Occam's razor recommends the simplest
assumption which does not introduce any new entity nor contradict any
known fact. Since exotic (e$^+$e$^-$) quasi-bound states are unable to
provide several excited states spaced at only a few hundred kilovolts,
as the EPOS' line separations demand, Occam's choice falls upon the
Quadronium (e$^+$e$^+$e$^-$e$^-$) atom. Although its binding mechanism
is inexplicable within our present knowledge, it is also not possible
at present to prove that it has no tightly bound states. For these
reasons, we have utilized the Q$_\circ$ atom, and the molecular states
it might form with nuclei, as a conceptual structure to organize the
the data of the Sharp Lepton Problem, and its discussion.

One day, perhaps, some data will be measured which could rule it out.
Then hopefully those data will point us towards a new direction. Or
perhaps one day we will begin to believe that the data really speaks in
favor of the existence of Q$_\circ$.  The we can turn to the difficult, and perhaps profoundly fundamental, problem of describing it
within the theory of Quantum Electrodynamics.

In the interim, we can ask whether the very existence of Q$_\circ$
might already imply some contradiction, not of nuclear data, but of the
data of QED. As a first step, we have inquired\cite{grif/95e} what
implications follow from the alteration of the four lepton
(e$^+$e$^+$e$^-$e$^-$) spectrum in QED by the requirement that it
exhibits one or more strongly bound state poles. The answer is that
Q$_\circ$'s strongest effect  upon contemporary QED would lie in its
modification of the decay rate of triplet orthopositronium\footnote{And
not in the value of (g$_e$-2), upon which  the X$_\circ$
particle\cite{rein/86e,rein/87}, which was assumed to couple directly
to (e$^+$e$^-$), was shown to have an intolerably large effect, but
which Q$_\circ$ influences only in order $\alpha^4$, smaller than the
current experimental uncertainty\cite{grif/95e}.},  a quantity already
remarkable for the persistent 10$\sigma$
difference\cite{mill/90e,kino/90a} between its measured value and the
best available calculation.

\subsection{Sharp Annihilative Positron Emission}

One other still untested experimental implication of the Bound
Annihilative Pair Decay in the Composite Particle Scenario is the decay
by Sharp Annihilative Positron Emission\cite{grif/91a,grif/89c} in
which\footnote{In addition, one photon emission, in which the full
$\sim$1.8 MeV energy of the Q$_o$ is emitted as a single photon, and
the more elusive  (``bipositron'') emission of $(e^+e^+)$ and
(``tri-lepton'') emission of $(e^+e^+e^-)$, can
follow\cite{grif/94,grif/91c} Q$_o$ decay in a bound
\{Z,Q$_o$\}state.}, in which the electron of a decay pair from a
bound \{Z,Q$_o$\} molecule is captured into an unoccupied Bohr orbit in
the heavy nucleus, Z. Then only the positron emerges, carrying the
total decay kinetic energy plus the added binding energy difference
between the molecule and the bound electron. A bound 805 keV pair decay
line would emit a series of such annihilative positrons with  sharp (in
the nucleus' rest frame) kinetic energies\footnote{Ref.\cite{grif/91a}
discusses these lines in more detail and estimates their relative
emission probabilities.} of about 805+132=937, 838,
820,...keV\footnote{specified by the Bohr electron binding energies,
where the 1s binding energy has been taken to be 132 keV.}.  The
occurrence of such positrons would provide strong evidence that the
source of the sharp pairs is a composite particle which can bind to a
high-Z nucleus.

\section{Summary and Conclusions}

The published APEX data exhibit a sharp pair line (123$\pm$46 counts)
of width 23.4 keV at a sum energy of 793$\pm$7 keV, according to a
chi-squared analysis.  Confidence Level analysis of the APEX data also
implies at the 99\% confidence level a mean sharp pair value greater
than 23 and fewer than 146 sharp pairs in the 20 keV bin at 790 keV.
For our best fitting line shape, these bounds correspond to total sharp
pair counts greater than 24 and less than 217. These results, and
others based upon the combining of two and three bins, all agree that
some sharp pair count greater than 23 and less than 252 should be
expected in any repetition of the APEX experiment. All of these
implications of the APEX data also honor the sharp pair upper bound
(292 counts near 800 keV) inferred by APEX from the same data.

Regarding the question of conflict between the APEX and EPOS
experiments, we have presented quantitative purely empirical indices of the two data
bases which show that, although  APEX' measured luminosity (as inferred
from the total positron counts and efficiencies of the two experiments)
is $\approx3\times$ larger than EPOS' , nevertheless  APEX' actually counted
altogether 20\% fewer (40.8K/50K) ``Qualitatively Similar''  EPOS-type
pairs than were counted by the EPOS experiment, and some 20\% more
(1480/1280) pairs per 20 keV near 800 keV.  These ratios show that for
direct APEX/EPOS comparisons of coincident pairs, the APEX data base is
rather comparable to the EPOS', and certainly not substantially
larger.  Furthermore, the ratios of sharp pairs to background pairs
near 800 keV agree within 10\% for the two experiments, suggesting
mutual consistency between APEX and EPOS also with regard to their {\it
sharp} pairs near 800 keV.

Thus, under purely empirical criteria, the APEX data
must be judged to be at least weakly corroborative of the EPOS' 800 keV
sharp line. On the other hand, the APEX evidence is very comparable with EPOS:
no one who doubted the EPOS results needs to be convinced by those of APEX.

Consideration of the APEX collaboration's {\it expectations} for sharp
pairs vis a vis EPOS' data indicates that those expectations inflate
the implications of the EPOS data arbitrarily  and should therefore be
disregarded in assessing the experimental situation.

A puzzle emerges why, despite  the fact that its luminosity is
($\approx3.2\times$) larger than EPOS, APEX measures roughly the same number,
both of background and of sharp EPOS-type RL(1,1) pairs, as does EPOS.
This means that the APEX' Gathering Power\footnote{The ``Gathering
Power'' for a certain emission process (defined explicitly for sharp
pairs in Table V) of a detecting apparatus specifies the fraction of
the emissions created at the target which are actually counted by the
apparatus.} for  RL(1,1) pairs is not comparable to EPOS', as one might
have expected, but $\sim2\times$ to $3\times$ smaller. The pair data allows us to
estimate specifically that EPOS apparent detection efficiency for
background RL(1,1) pairs is about $3.3\times$ larger than APEX, and that its
apparent efficiency for sharp pairs near 800 keV is at least $2.2\times$
larger.  The latter result, if taken literally, flatly contradicts the
ratio of the published values, for reasons not yet understood. But this
discrepancy might speak instead to the possibility  that the (1,1) pair
events occur more probably in the front layer of the target, where the
projectile ionization charge is significantly larger that its thick
target equilibrium value. In such a scenario, it is the effective
values of the luminosities which would require adjustment, rather than
the pair detection efficiencies.

In short, the APEX experiment provides independent corroborative
evidence for a sharp (e$^+$e$^-$) pair line near 800 keV. Furthermore,
the background RL(1,1) pair distributions of the APEX and EPOS
experiments are roughly comparable in size and shape. In that context,
the rough similarity also of their sharp pair counts can be considered
as merely another aspect of an overall consistency between their
distributions. Thus by purely empirical measures, APEX' and EPOS' pair
results agree, although APEX' RL(1,1) pair counts are $\sim3\times$ fewer
than expected from EPOS' and the luminosities. In the end there is no
purely empirical basis in the APEX experiment to support the claim that
it contradicts the EPOS' sharp pair results.

A brief perspective review emphasizes the large body of evidence which
has accumulated over the fourteen years of research into what has come
to be the Sharp Lepton Problem. Non-Heavy-Ion alternative studies of
few MeV positrons (or electrons) and of $\sim$1.8 MeV gamma rays
incident upon high-Z atoms to yield sharp pairs, and a search for Sharp
Annihilative Positrons in the 600 to 900 keV range are recommended as
independent and perhaps more easily repeatable avenues of
investigation.

\section{Acknowledgement}
The  author is grateful to Drs. Thomas Cowan, Mark Wolanski, K.C. Felix Chan,
Frank Wolfs, and Thomas Trainor for their help in the course of this work.
The support of the U. S. Department of
Energy under grant no. DE-FG02-93ER-40762 is gratefully acknowledged.\

\section{Figure Captions}

Figure 1 displays the total coincident pair counts, $N_{OBS}$, the
event-mixed background distribution, $N_{BK}$, and the APEX'simulation
calculation of the sharp pair energy distribution (multiplied by 0.1)
from the decay of a source created with a cross section of
5.0$\mu$b/sr. These data were first published in  Figure 2a of  the
APEX report\cite{ahma/95}.\\

\vspace{0.05in}
Figure 2 presents the data from Figure 1 in the range from 500 to 920
keV where previous experiments reported evidence for sharp pairs. The
largest excess (3.2$\sigma$) above the background occurs at 790 keV.
The best gaussian fit near 790 keV produces the cross-hatched addition
to the background, and is described in the text. The data are tabulated
in Table I.\\

\vspace{0.05in}
Figure 3 presents the number of counts in excess of the background,
$N_{EXC}$, and plots (crosshatched) the best gauss fit near 800 keV.
Also plotted are the 99\% CL lower (24.2) and upper (146) bounds upon
the mean excess pair count in the 20 keV bin at 790  keV. For the
best-fitting  line shape these bounds define a total line strength
between 35 and 217.  The bounds for all twenty one bins are listed in
Table I.  Except for the 790  keV bin, all the lower bounds are
negligible ($\leq$2), and all of the the upper bounds are less than
100($\approx$3.8$\sigma$).\\

\vspace{0.05in}
Figure 4. The normalized  APEX and EPOS luminosity densities are
plotted versus the beam energy. APEX' single thick target run provides
a uniform luminosity density of 5.9\% per 0.01 MeV/U  whereas EPOS five
thin target runs give the beam energy dependent luminosity shown. EPOS'
809 keV sharp  pairs were seen only in their two highest energy runs,
which covered the 0.07 (MeV/U) intervals up to 5.87 (Mev/U) and 5.90
(MeV/U). The fractional luminosity density in the 0.02 MeV/U interval
common to these runs but not covered by any of the other three runs,
 where a narrowly resonant pair production cross section could have
contributed pairs to the two runs in which they were observed but to no
others, is 5.7\% per 0.01 MeV/U.\\

%\bibliography{/homes/ntg/griffin/ltex/biblio} 
%\bibliographystyle{prsty}


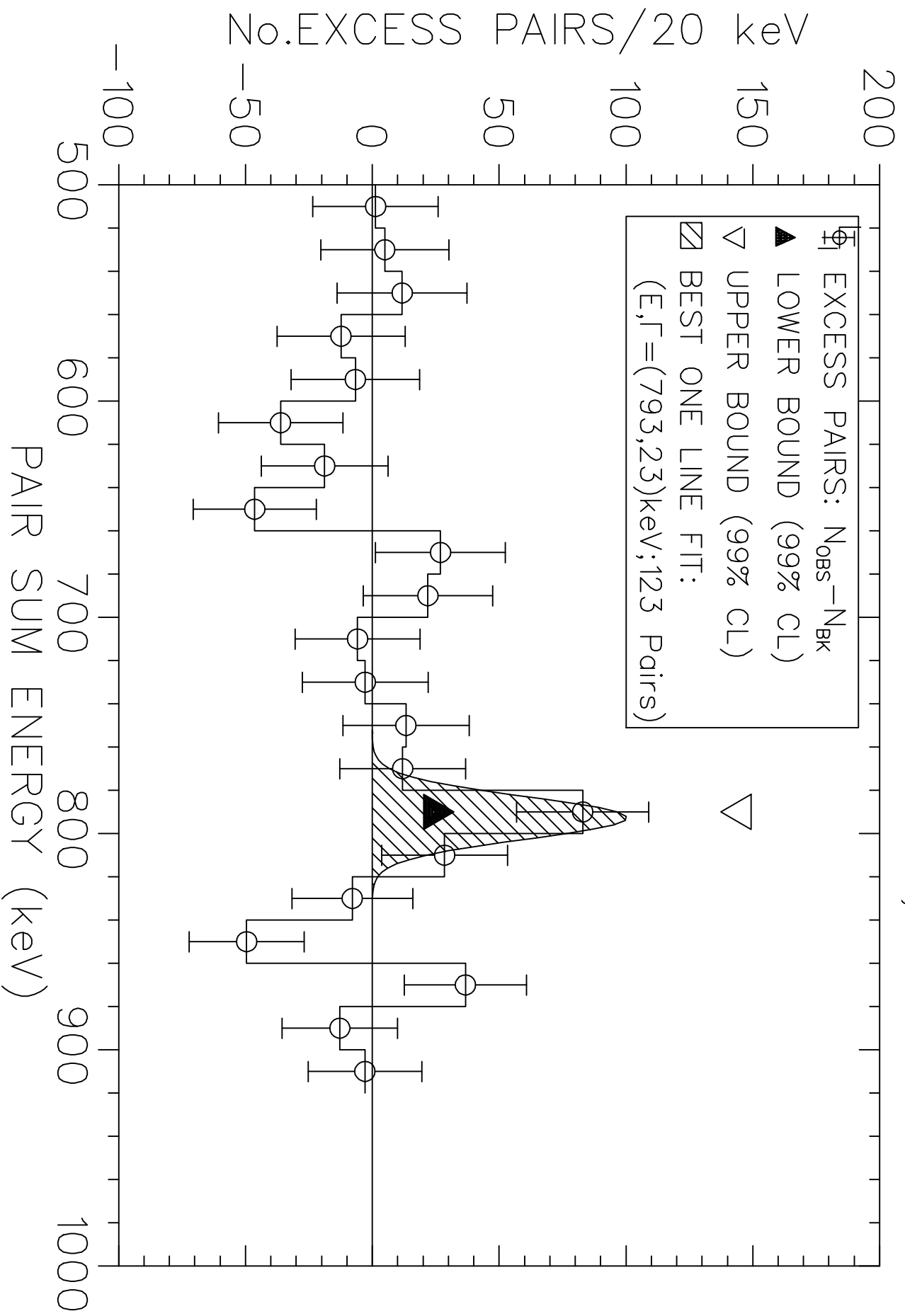
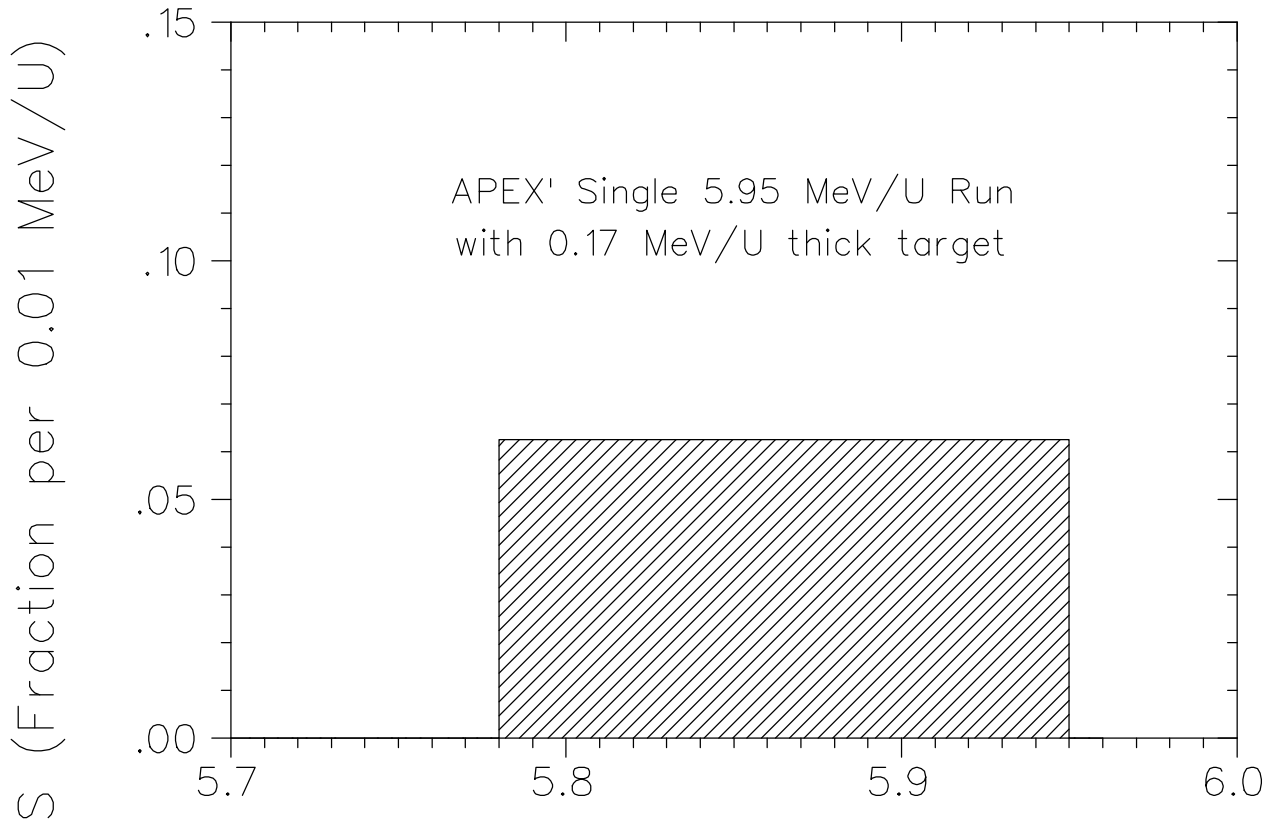
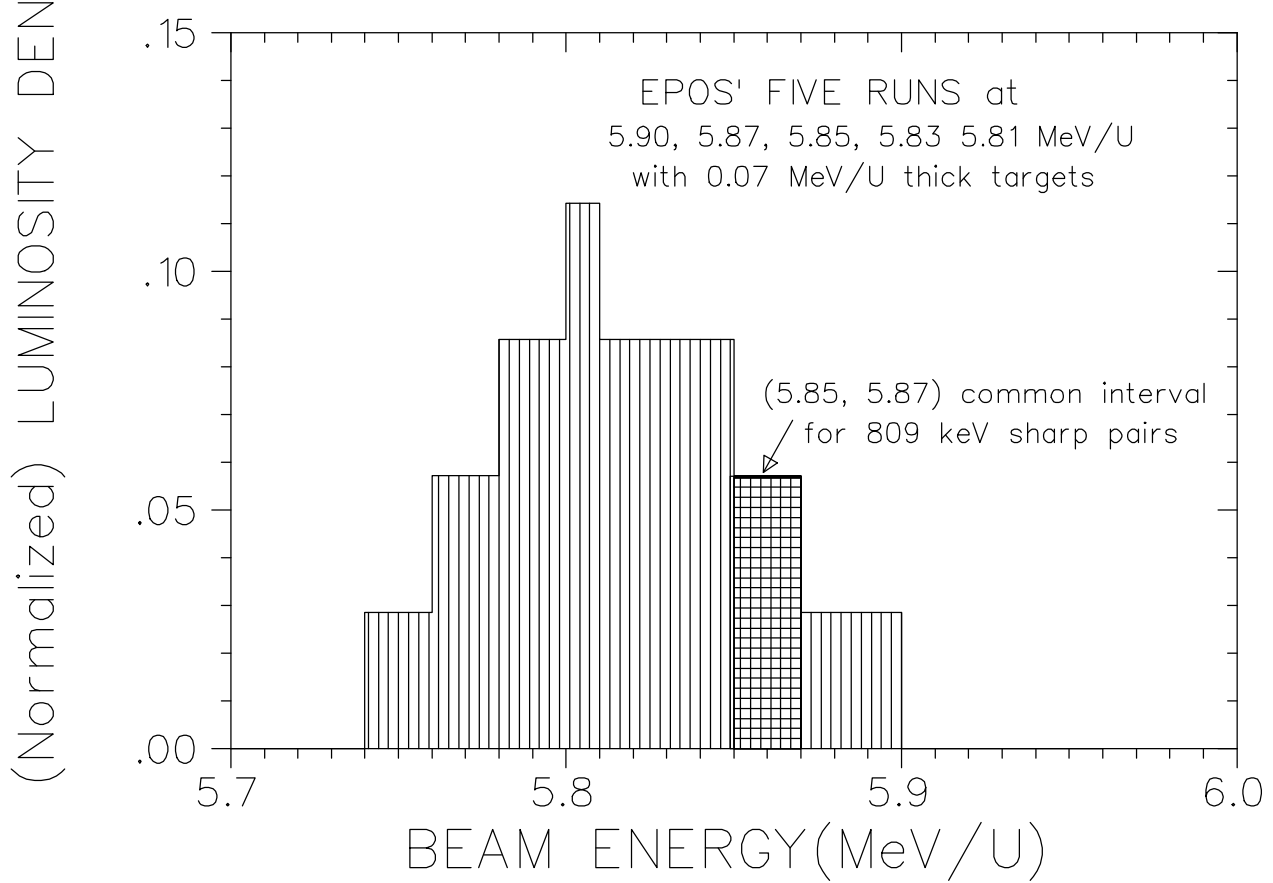
\end{document}